\documentclass[11pt,twocolumn]{aastex701}

\usepackage[utf8]{inputenc}
\usepackage[english]{babel}
\usepackage{subfig}
\usepackage{caption}
\usepackage{graphicx}
\usepackage{float}
\usepackage{lineno}
\usepackage{subcaption}
\usepackage{amsmath}
\usepackage{booktabs}
\DeclareUnicodeCharacter{2212}{-}
\usepackage{mwe}

\begin{document}

\title{Searching for Extragalactic Exoplanets: A Survey of the Sagittarius Dwarf Galaxy Stream with TESS}

\author[0000-0001-5255-8274]{William Schap}
\affiliation{Department of Astronomy, University of Florida, P.O. Box 112055, Gainesville, FL, USA}
\email{wschap@ufl.edu}

\author[0000-0001-7730-2240]{Jason Dittmann}
\affiliation{Department of Astronomy, University of Florida, P.O. Box 112055, Gainesville, FL, USA}
\email{jasondittmann@ufl.edu}

\author{Elizabeth Lada}
\affiliation{Department of Astronomy, University of Florida, P.O. Box 112055, Gainesville, FL, USA}
\email{elada@ufl.edu}

\accepted{for publication in AJ February 09, 2026}

\begin{abstract}

To date no exoplanets have been detected outside the Milky Way, and their extragalactic occurrence rates are poorly constrained. Using available data from TESS we perform the first transit survey of the Sagittarius dwarf galaxy stream using 15,176 main sequence stars identified as likely members. We calculate an upper limit of $<$1.01\% for hot Jupiters with radii of 1-2 R$_{Jup}$ and periods of 0.6-10 days after detecting zero planets. We compare our calculated occurrence rate upper limits to the upper limits found in the Milky Way globular clusters M4 and 47 Tuc. Our 1-$\sigma$ occurrence rate upper limit of $<$0.37\% for the Sagittarius dwarf galaxy stream, for planets with radii of 1.5-2 R$_{Jup}$ and periods $<$10 days, is lower than the $<$0.57\% upper limit measured in 47 Tuc. Similarly, our 2 sigma upper limit of $<$0.78\% for planets with radii of 1.4-2 $_{Jup}$ and periods $<$8 days is below the $<$0.81\% upper limit measured in M4. {{We predict that a future analysis of TESS data with a high detection efficiency for hot Jupiter transit depths would require $\eta_{extragalactic}$ $\geq$ 11,467 target stars to detect a planet of extragalactic origin. Therefore, we predict that a future investigation of TESS data which includes additional extragalactic stellar streams will be able to either detect the first extragalactic origin planet or provide evidence that older, lower metallicity extragalactic environments may experience a lower hot Jupiter occurrence rate than is observed for the Milky Way.}}

\end{abstract}

\section{\label{sec:intro}Introduction}

Since the discovery of the first hot Jupiter in 1995 \citep{MayorandQueloz1995}, large transit surveys performed using Kepler and TESS have provided constraints on their frequency. These surveys have observed occurrence rates of 0.4-0.6\% across the Milky Way \citep{Howard2012ApJS..201...15H,Fressin2013ApJ...766...81F, Mulders2015ApJ...798..112M, Petigura2018AJ....155...89P, Kunimoto2020AJ....159..248K}, which have been shown to vary with different host star properties such as stellar mass \citep{Wright2012ApJ...753..160W,Bayliss2011ApJ...743..103B,gan2023,Bryant2023} and metallicity \citep{fischer2005,Boley2021AJ....162...85B,Thorngren2016ApJ...831...64T}. For extragalactic environments, this occurrence rate has yet to be constrained. The stellar populations of even the closest dwarf satellite galaxies are too dim for large scale transit surveys like TESS which focus on bright, nearby stars. However, the merger history of the Milky Way has produced streams of extragalactic stars through the tidal disruption of satellite dwarf galaxies that provide extragalactic sources within the Milky Way halo \citep{Johnston1996ApJ...465..278J}. Using the precise photometry and kinematic properties of halo stars made available by Gaia, over 20 different Galactic Halo streams have now been identified in the Milky Way halo \citep{Grillmair2016ASSL..420...87G} including Helmi \citep{Helmi1999Natur.402...53H}, Sagittarius \citep{Ibata1994Natur.370..194I}, and Gaia-Enceladus \citep{Helmi2018Natur.563...85H}.

Although these stellar streams provide populations of extragalactic origin stars that are much closer than the stars in the Milky Way satellite galaxies, they are much dimmer than the nearby bright stars the Kepler  \citep{Borucki2010Sci...327..977B} and TESS \citep{ricker2015} missions were designed to survey. This has limited any potential large scale transit surveys of individual extragalactic origin stellar streams using currently available Kepler or TESS data. However, the recent release of Gaia DR3 and the light curve reduction software packages \texttt{eleanor} and \texttt{TESS-Gaia Light Curve (TGLC)} have provided the opportunity to produce light curves for stars in the TESS Full Frame Images that are much dimmer than the targeted magnitude of the TESS mission. These packages have been used to produce millions of precomputed light curves for stars in the TESS Full Frame Images as dim as T$_{mag}$=16, and allow users to produce new light curves for even dimmer stars \citep{Feinstein2019PASP..131i4502F,han2023}. The ability to create corrected light curves for dim stars makes a transit search of extragalactic stellar streams using the existing TESS data possible.  

{{In \citet{ramos22}, they identify a population of 773,612 halo stars as candidate members of the Sagittarius dwarf galaxy stream. This sample of potential members includes 51,208 stars brighter than T$_{mag}$=16, and 138,343 stars brighter than T$_{mag}$=17.}} We conduct a transit survey using the brightest stars from this catalog to search for hot Jupiters and provide the first constraints on their occurrence rate in the extragalactic environment of the Sagittarius dwarf galaxy stream. We define our target list of Sagittarius dwarf galaxy stream stars in Section \ref{sec:methods} and introduce the \texttt{eleanor} and \texttt{TGLC} python packages used to produce reduced light curves for our dim star targets. After performing a Box Least Squares (BLS) periodogram analysis on our reduced light curves, we vet the strongest candidate from our survey in Section \ref{sec:candidates}. In \ref{sec:recovery} we present our injection recovery results and show our surveys completeness to hot Jupiters with radii of 1-2$R_{Jup}$ and periods less than 10 days. In Section \ref{sec:discussion} we place upper limits on the occurrence rate of hot Jupiters in the Sagittarius dwarf galaxy stream, and compare our upper limits to star forming environments in the Milky Way halo. Finally, we summarize our results and conclusions in Section \ref{sec:conclusion}.

\section{\label{sec:methods}Methods}

\subsection{\label{sec:targetlist}Creation of the Sagittarius Dwarf Galaxy Stream Target List}
We created our target list using a subset of the 773,612 stars classified by \citet{ramos22} as potential members of the Sagittarius dwarf galaxy stream with magnitudes brighter than G=19.75. These stars were identified by selecting halo stars with Gaia eDR3 proper motions that follow the known kinematic substructures present in the Sagittarius dwarf galaxy stream. To calculate the likelihood of a given star belonging to a specific branch of the Sagittarius dwarf galaxy stream, they compared the on-sky position of each Gaia identified member to models of the known branches.

We cross matched these stars from \citet{ramos22} with the TESS Input catalog (TIC) \citep{Stassun2019AJ....158..138S} to obtain stellar parameters for each star such as TIC ID, TESS magnitude (T$_{mag}$), luminosity, T$_{eff}$, radius, and mass. From this list we then selected main sequence stars suitable for our survey using the following criteria:
\begin{enumerate}
  \item Stars brighter than T$_{mag}$$\leq$17 based on the photometric precision predictions for \texttt{eleanor} \citep{Feinstein2019PASP..131i4502F} and \texttt{TGLC} \citep{han2023}.
  \item Stars calculated to have Sagittarius dwarf galaxy membership likelihoods of $>$50\% \citep{ramos22}.
  \item Stars with stellar radii $<$2$R_\odot$ to focus on the types of main sequence FGK stars primarily surveyed by Kepler and TESS.

\end{enumerate}

{{Our initial cutoff selected the 138,343 stars from \citet{ramos22} that had T$_{mag}$ brighter than 17. We then chose to select only stars which had membership probabilities of $>$50\%. In \citet{ramos22} they determine this membership probability for their sample by first searching for kinematic over-densities, and using the available data on the stars in their sample to determine which over-densities could be attributed to the Sagittarius dwarf galaxy. After separating out the kinematic substructure of the Sagittarius dwarf galaxy stream, they modeled the 4 branches of the stream as Gaussians with the same angular size along the branch. Finally based on a star's sky position the probability of membership was determined using the equation: }}

\begin{equation}
\begin{split}
   \textnormal{Prob}(\textnormal{Sgr}|\widetilde{\Lambda}_{\odot},\widetilde{\beta}_{\odot}) = \textnormal{Prob(A} | \widetilde{\Lambda}_{\odot}, \widetilde{\beta}_{\odot}) + \textnormal{Prob(B}|\widetilde{\Lambda}_{\odot}, \widetilde{\beta}_{\odot})\\
 −\textnormal{Prob(A} | \widetilde{\Lambda}_{\odot}, \widetilde{\beta}_{\odot})\textnormal{Prob(B}| \widetilde{\Lambda}_{\odot}, \widetilde{\beta}_{\odot}),
\label{eqn:ramosprob}
\end{split}
\end{equation}

{{where $\widetilde{\Lambda}$ and $\widetilde{\beta}$ are the spherical and angular coordinates along the Sagittarius dwarf galaxy stream, and $\textnormal{Prob(}A|\widetilde{\Lambda}_{\odot}, \widetilde{\beta}_{\odot})$ and $\textnormal{Prob(B}|\widetilde{\Lambda}_{\odot}, \widetilde{\beta}_{\odot})$ are the normalized Gaussian probabilities of belonging to the bright and faint branches of the stream. Using these calculated probabilities, we selected the 114,448 stars in our sample with membership probabilities $>$50\%. Finally, we selected the subset of stars in our sample with stellar radii of $<$2$R_\odot$, which resulted in a final candidate list of 20,205 stars.}}

This list included stars dimmer than the T$_{mag}$ = 13.5 cutoff of the TESS Quick-Look pipeline \citep{Huang2020RNAAS...4..204H} and TESS Science Processing Operations Center (SPOC) pipeline \citep{Caldwell2020RNAAS...4..201C}, which are the standard light curve reduction software packages for TESS data. Due to this magnitude cutoff we instead reduced our light curves using the software packages \texttt{eleanor} \citep{Feinstein2019PASP..131i4502F} and \texttt{TGLC} \citep{han2023} which are able to create accurate light curves for stars down to our T$_{mag}$ = 17 cutoff. {{We chose to use both \texttt{eleanor} and \texttt{TGLC} because each program reduces light curves using a slightly different method (see Section \ref{sec:reductionsoftware}), and this allowed us to select the best possible reduction for the final set of light curves used in our analysis.}}

{{One of those differences is how each deals with dilution correction. While \texttt{eleanor} primarily depends on variable apertures and background subtraction to account for dilution, \texttt{TGLC} makes use of PSF removal using data from Gaia for each star (see Section \ref{sec:reductionsoftware} below). A recent study using \texttt{TGLC} found that planets detected with the various publicly available TESS photometry pipelines may at times underestimate planet radii by a weighted median of 6.1\% $\pm$ 0.3\% \citep{Han2025ApJ...988L...4H}. Because of this and the difference in dilution correction, we saved the light curves created by \texttt{eleanor} and \texttt{TGLC} so that any candidate signals found from our chosen method could be cross checked with the result from the other method in order to reject large companions that may be misidentified as planetary sized objects with radii of $<$2 R$_{Jup}$.}}

\subsection{\label{sec:reductionsoftware} \texttt{eleanor} and \texttt{TGLC}}

Our analysis pipeline first creates a light curve using \texttt{eleanor} \citep{Feinstein2019PASP..131i4502F}, which was designed to produce corrected light curves for stars within the TESS Full Frame Images with magnitudes brighter than I=16. \texttt{eleanor} begins by correcting for the offsets in RA and Dec for each star in each Full Frame Image with 7.5 $\leq$ T$_{mag}$ $\leq$ 12.5 by comparing them to the expected RA and DEC values reported by the TESS Input Catalog. A pointing correction is applied to the entire Full Frame Image which minimizes the square of the differences in sky position between the observed and expected coordinates, and a time-stacked and background subtracted 148 × 104 pixel cutout (referred to as a ``postcard") is created. Background subtraction is performed on the ``postcard" using the \texttt{photutils} function \texttt{MMMBackground}, and a library of aperture shapes and sizes are then tested to determine which aperture minimizes the root mean square of the photometric noise. 

To minimize stray light contamination for stars found in crowded fields, \texttt{eleanor} provides a crowded field option which restricts its aperture selection testing to only the smallest apertures in its aperture library. We select the crowded field option by setting \texttt{aperture\_mode} to ``small" for our analysis due to the crowded nature of the Sagittarius dwarf galaxy stream, and set \texttt{sectors} to ``all" in order to produce multi-sector light curves for each star using all available sectors per star. 

We further set the height and width values to 15 pixels, and the background size to 31 pixels. {{We set \texttt{do\_pca} to ``True" to allow \texttt{eleanor} to use the cotrending basis vectors on the background subtracted light curve. With \texttt{regressors} set to ``corner" \texttt{eleanor} uses the four pixels in the corners of the postcard to infer the background. We chose to set \texttt{do\_psf} to ``False", as \texttt{eleanor} \citep{Feinstein2019PASP..131i4502F} identifies this task as having known issues for faint stars (I $\geq$ $\sim$15). After \texttt{eleanor} finished running, quality-flagged photometry points were removed from the light curve. These quality flags are reported by TESS to note any issues that occurred during observation that impacted quality of the photometry during that time period.}}

Following the creation of the \texttt{eleanor} based light curve, we then reduce the same target star using \texttt{TGLC} \citep{han2023}. \texttt{TGLC} is designed to incorporate Gaia DR3 data in order to subtract the PSFs of nearby contaminant stars from the TESS Full Frame Image cutouts and reach a reported photometric precision of $<$2\% for T$_{mag}$=16 stars in 30 minute data \citep{han2023}. To do this \texttt{TGLC} first creates a Full Frame Image cutout around the chosen target star. It then converts the Gaia positions of each star in each Full Frame Image into pixel positions by propagating the Gaia positions to the median TESS epoch, and then converting the propagated RA and DEC values using the world coordinate system (WCS) information from the Full Frame Image header. \texttt{TGLC} then creates a simulated Full Frame Image by forward modeling the best-fit PSF shapes for all stars in the Full Frame image except for the target star, and then subtracting the simulated image from the observed Full Frame Image data. This results in a decontaminated residual image of the target star. 

The final effective PSF (ePSF) model for the target star is then created by simultaneously fitting the ePSF and background model parameters in the decontaminated image. From these calibrated cutouts, both PSF and aperture photometry light curves are created and stored in the final \texttt{TGLC} data product. 
In addition to PSF and aperture photometry, the authors note in \cite{han2023} that in some fields, a weighted average of 0.4 PSF + 0.6 Aperture photometry shows an increase in precision when compared to each of the photometry methods individually. We therefore create light curves using aperture photometry, PSF photometry, and a weighted average of 0.4 PSF + 0.6 aperture photometry to determine the optimal photometry method for each star in our target list.

For our analysis we chose to run \texttt{TGLC} using a Full Frame Image cutout size of 50 pixels per star as larger sized cutouts did not provide significant increases in precision for our dataset. We set a TESS magnitude upper limit of 17 for non-target stars processed as part of the data reduction, as it removes any contaminant stars down to the magnitude limit of our target list, and is the magnitude where the resulting increase in precision began to plateau due to dimmer stars beginning to blend with the Full Frame Image background. {{We set \texttt{sector} to ``None" and ``\texttt{first\_sector\_only}" to ``False" as this combination produces a multi-sector analysis for a given TIC ID. \texttt{prior} is set to ``None" which does not allow the field stars to float, and is the setting suggested by the authors for an initial light curve reduction. }}Once \texttt{TGLC} finished running, we removed all quality-flagged photometry points from the final light curves.

\subsection{\label{sec:additionalcorrections} Additional Light Curve Corrections and Refinement of Target Sample}

{{While \texttt{eleanor} and \texttt{TGLC} are both designed to provide accurate light curves for faint stars, our target list pushes the magnitude limits of even these software packages. During the reduction of our multisector data this meant that there were still occasions where individual sectors would have scatter levels that were too high to be useful to our analysis and needed to be removed. To ensure that we were not excluding possible transit detections with our chosen precision cutoffs we performed injection recovery tests where we modeled a series of light curves with varying levels of photometric scatter and determined the level at which injected signals of 2 R$_{Jup}$ and below became undetectable, and varied based on the magnitude and radius of each star. By determining this light curve precision cutoff we were confident that we were only rejecting light curves where the scatter was high enough that planet sized objects were no longer detectable.}}

To remove any remaining noise spikes that remained while preserving any small candidate signals that may exist, we then removed only data points that appeared 5$\sigma$ or more above the median of the normalized light curves produced by \texttt{eleanor} and \texttt{TGLC}. Following the removal of this noisy data, we binned each sector down to a 30 minute cadence to create consistency across our final multi-sector light curves. 

\begin{figure}
\centering
\includegraphics[width=8.0cm]{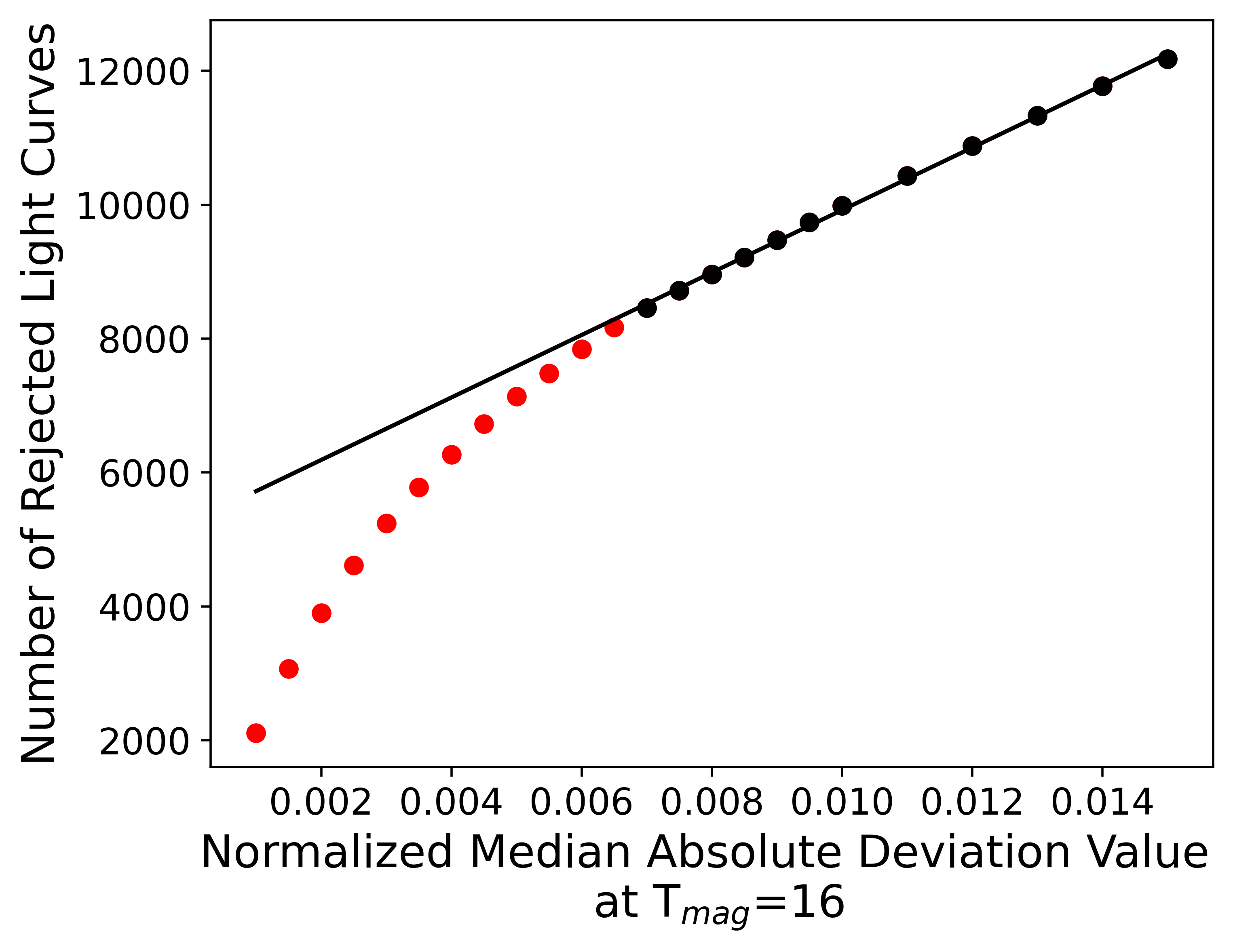}
\caption{Comparison between the selected T$_{mag}$=16 normalized median absolute deviation value used as an anchor point for our Poisson-like noise cutoff, and the number of rejected stars at that noise cutoff. At a normalized median absolute deviation anchor point of 0.0065 the function becomes linear out to brighter median absolute deviation cutoffs (shown in black). Due to significant contamination from nearby bright neighboring stars the normalized median absolute deviation values of some target stars fall below the expected noise limit. The transition at 0.0065 represents the point where our noise cutoff removes these poorly fit light curves.} 
\label{cutoff}
\end{figure}

Significant contamination from nearby bright stars in crowded regions can lead to light curves with scatter values that fall below the predicted noise limit \citep{Feinstein2019PASP..131i4502F}. We removed these poorly reduced light curves by selecting a Poisson-like noise threshold and rejecting any light curves with a median absolute deviation flux value that fell below our selected noise threshold. We start with the Poisson noise in magnitudes as 
\begin{equation}
  \sigma_{m} = \frac{2.5}{ln10}*\frac{\sqrt{F}}{F},
\label{eqn:poisson}
\end{equation}
where the flux (F) can be defined as a combination of magnitude (m) and a constant (C) using
\begin{equation}
  F = 10^{-\frac{m-C}{2.5}}
\label{eqn:fluxmag}
\end{equation}
to produce the equation
\begin{equation}
  \sigma_{m} = \frac{2.5}{ln10}*\frac{\sqrt{10^{-\frac{m-C}{2.5}}}}{10^{-\frac{m-C}{2.5}}}.
\label{eqn:poissonmag}
\end{equation}
We determined the scaling of our noise threshold by anchoring the function to a series of $\sigma_{m}$ values at T$_{mag}$=16, and determining how many light curves were rejected at each anchor point. We selected T$_{mag}$=16 as it represents the brighter end of our survey's magnitude range, and \texttt{TGLC}'s reported photometric precision of $<$2\% for T$_{mag}$=16 stars in the TESS 30 minute cadence data \citet{han2023}. We compared the chosen anchor point values to the number of rejected light curves in Figure \ref{cutoff} and determined where the relationship became non-linear. This occurred at the point where the noise threshold transitioned from rejecting the poorly reduced light curves to rejecting the well fit light curves. We determined that this transition occurred at an anchor point value where the normalized median absolute deviation was 0.0065 at T$_{mag}$=16. For some of the stars in our target list both the \texttt{TGLC} and \texttt{eleanor} light curves fell above this noise threshold, and we selected the light curve with the highest precision for our analysis. For stars where only one of the reduced light curves fell below our threshold, we selected the remaining light curve for our analysis. For stars where both light curves fell below the threshold, the star was excluded from our target list. Our final target list consisted of 15,176 stars with a magnitude range of T$_{mag}$ = 15.29-17. 
{{For each star's light curve we quantify how much our scatter deviates from the median value by determining the normalized median absolute deviation using the following equation:
\begin{equation}
  MAD = median(abs(F_i-F_{med})),
\label{scatterequation}
\end{equation}
where F$_i$ is an individual flux measure, and F$_{med}$ is the median flux value of the entire light curve. If the median absolute deviation of a given section was larger than the depth of a 2 R$_{Jup}$ signal, that sector was removed as our injection recovery testing showed that planet sized signals around those stars would not be able to be detected by our pipeline.}}

\begin{figure}
\centering
\includegraphics[width=8.5cm]{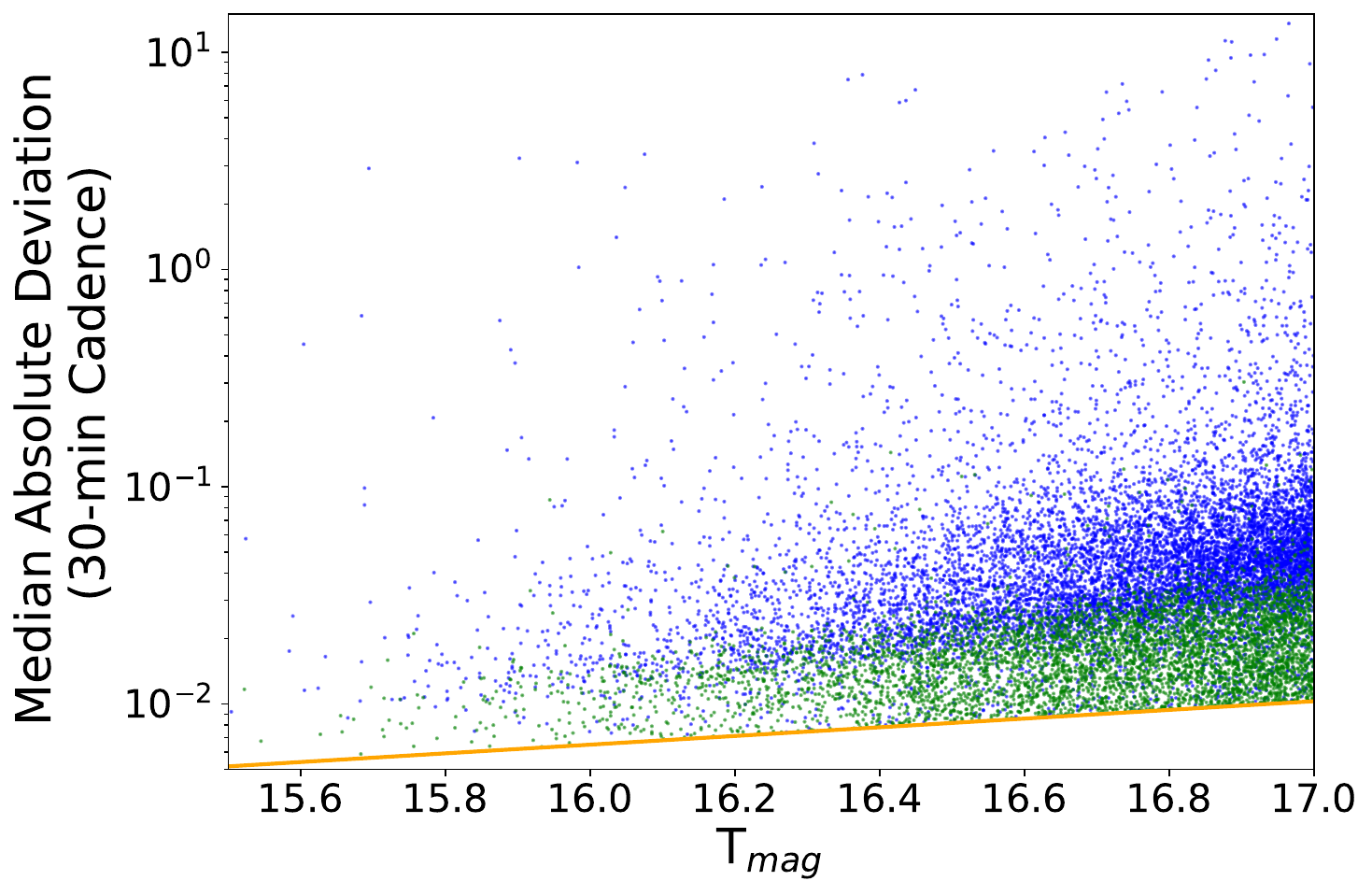}
\caption{The normalized median absolute deviation of each light curve in our survey as a function of T$_{mag}$ where the 9,768 light curves reduced with \texttt{TGLC} are shown blue, and the 5,408 light curves reduced with \texttt{eleanor} are shown in green. For the faint target stars stars observed in the most crowded fields of our survey, significant contamination from nearby bright neighboring stars can lead to light curves with median absolute deviation values that fall below the predicted noise limit (\citep{Feinstein2019PASP..131i4502F}. To remove these highly contaminated light curves we plotted the number of stars rejected at a series of Poisson-like noise cutoffs and selected the cutoff where the function began to deviate from the linear function seen at brighter median absolute deviation cutoffs (see Figure \ref{cutoff}). We show this selected noise cutoff as an orange curve in the scatter plot.}
\label{fig:scatter}
\end{figure}

We present the stellar parameters of each target star alongside the normalized median absolute deviation value of the selected light curve in Table \ref{tbl:stars}. In Figure \ref{fig:scatter} we plot our selected Poisson-like cutoff in orange alongside the T$_{mag}$ and median normalized median absolute deviation value of each light curve used in our analysis. For our target list of 15,176 stars in the Sagittarius dwarf galaxy stream, we show the 9,768 light curves reduced using \texttt{TGLC} in blue, and the 5,408 reduced using \texttt{eleanor} in green.

\begin{figure}[]
  \centering
  \begin{tabular}{@{}c@{}}
    \includegraphics[width=8cm]{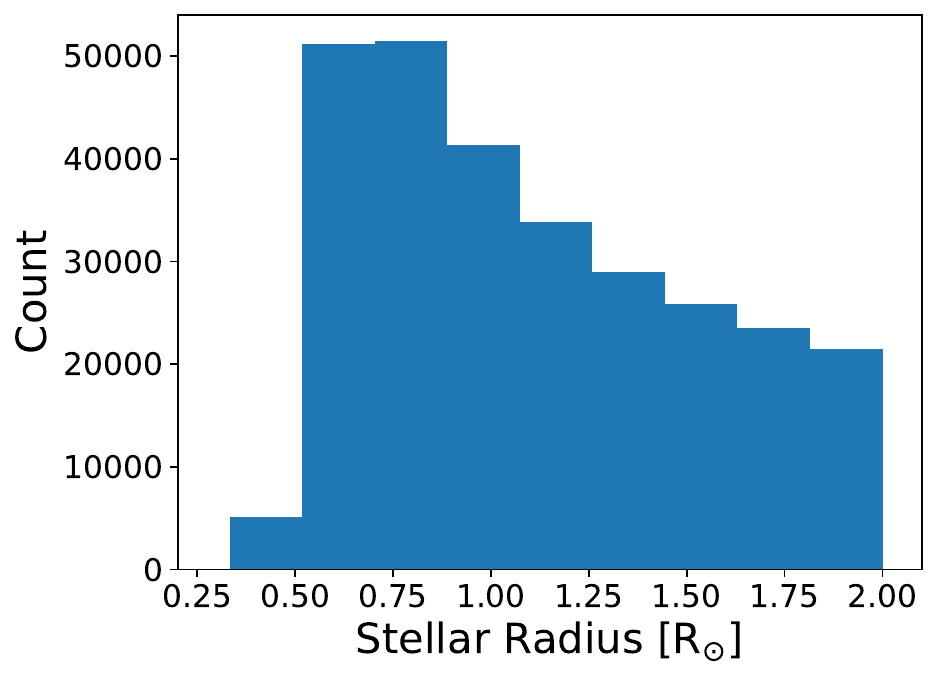} \\
    \small (a)
  \end{tabular}

  \begin{tabular}{@{}c@{}}
    \includegraphics[width=8.5cm]{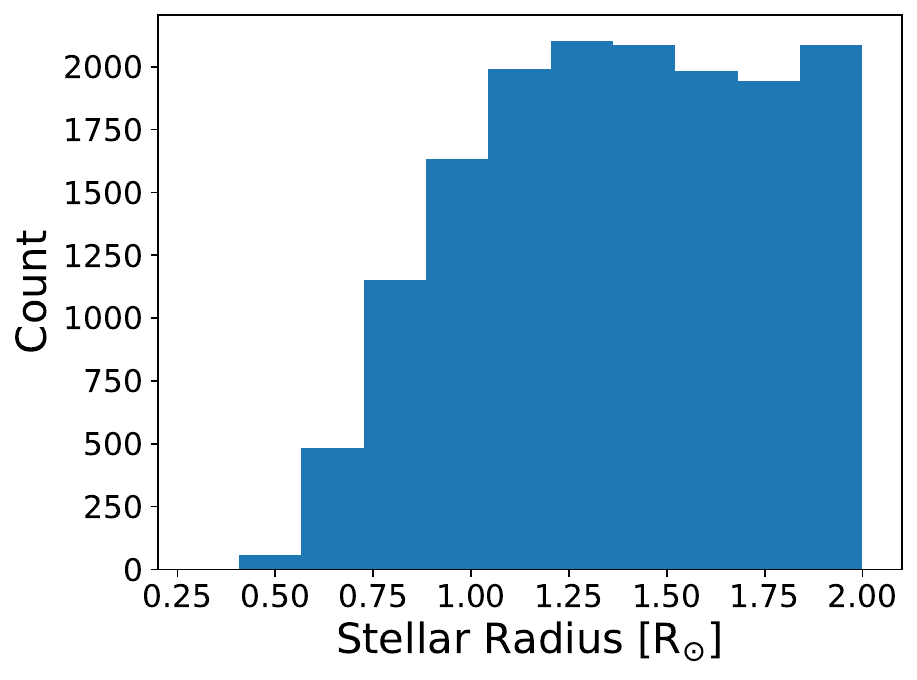} \\
    \small (b) 
  \end{tabular}

  \caption{(a) The stellar radius distribution of the stars from \cite{ramos22} with our cuts of R$_*$$<$2 R$_{\odot}$ and Sagittarius membership probability $>$50$\%$ across the full magnitude range of G $<$ 19.75. The median stellar radius of this distribution is 1.04 R$_\odot$. (b) The stellar radius distribution for the 15,176 stars in our survey with our cuts of R$_*$$<$2 R$_{\odot}$, Sagittarius membership probability $>$50$\%$, and T$_{mag}$$<$17. The median stellar radius of this distribution is 1.38 R$_{\odot}$.}
  \label{fig:radhist}
\end{figure}

\begin{figure}[]
\centering
\includegraphics[width=9.5cm]{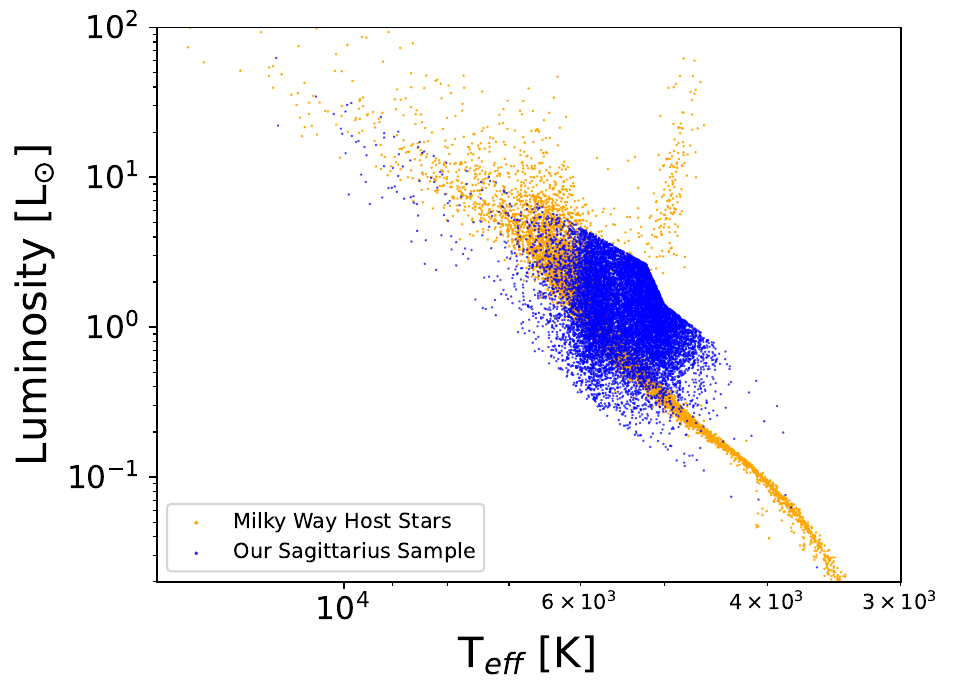}
\caption{{{HR diagram depicting our sample of 15,176 probable Sagittarius dwarf galaxy stream members in blue against the 7,993 planet host stars from the Kepler, K2, and TESS transit surveys \citep{Berger2023arXiv230111338B} in orange. The sharp cutoffs seen at higher luminosities are the result of magnitude limit cuts in the initial \cite{ramos22} stellar sample as well as our own cuts in stellar radius and T$_{mag}$. These cuts were used to reduce the number of evolved stars in our sample.}}}
\label{fig:HR}
\end{figure}

In Figure \ref{fig:radhist} we compare the stellar radius distribution of our target list to the full list of stars from \cite{ramos22} within the same range in stellar radius. Our target list contains a smaller fraction of low radii stars as all but the closest M dwarfs in the Sagittarius dwarf galaxy stream become fainter than our magnitude limit of T$_{mag}$$<$17. In Figure \ref{fig:HR} we show the HR diagram of our target list, where the magnitude limit and G$_{BP}$ - G$_{RP}$ cuts used in \cite{ramos22} as well as our own cuts in stellar radius and T$_{mag}$ result in the sharp features present.

\tabcolsep=0.05cm
\begin{table*}[t]

\caption{\em{Stellar Parameters of Source List}}
\scalebox{0.8}{\begin{tabular}{cccccccccccccccc}
\hline \noalign {}
\hline \noalign {\smallskip}
\hline \noalign {\smallskip}
RAJ2000$^a$ & DEJ2000$^a$ &  TIC$^a$ &  G$^b$ & eG$^b$ &  Tmag$^a$ & e$_{Tmag}$$^a$ & R$_*$$^a$ & eR$_*$$^a$ &  T$_{eff}$$^a$ &   eT$_{eff}$$^a$ &    Lum$^a$ & eLum$^a$ &  Distance$^b$ &   eDistance$^b$ & {{MAD}}\\
(deg)& (deg)&  & (mag)&(mag)&(mag)&(mag)&(R$_{\odot}$)&(R$_{\odot}$)&(K)& (K)&(L$_{\odot}$)& (L$_{\odot}$)& (pc)&(pc)&{{}}\\
\hline \noalign {\smallskip} 
359.990669 & -19.106041 & 114807719 & 17.504 & 0.001 & 16.939 &  0.007 &           1.323 &               - & 5021.0 &   126.0 & 1.002 &    - & 3425.27 & 828.75 &0.039\\
359.964848 & -23.488913 & 114806895 & 16.852 & 0.001 & 16.337 &  0.007 &           1.365 &               - & 5313.0 &   124.0 & 1.338 &    - & 2984.78 & 687.29 &0.012\\
359.961637 & -18.928439 & 114807753 & 17.529 & 0.001 & 16.938 &  0.007 &           0.965 &               - & 4910.0 &   126.0 & 0.488 &    - & 2366.69 & 650.12 &0.031\\
359.951631 & -21.549931 & 114807234 & 17.416 & 0.001 & 16.940 &  0.007 &           0.778 &               - & 5499.0 &   125.0 & 0.498 &    - & 2415.46 & 613.36 &0.027\\
359.922667 & -23.760307 & 114806859 & 17.151 & 0.011 & 16.845 &  0.013 &           0.774 &               - & 6580.0 &   204.0 & 1.012 &    - & 3075.68 & 808.15 &0.030\\
359.889039 & -23.681086 & 114806872 & 17.506 & 0.001 & 16.994 &  0.007 &           0.751 &               - & 5341.0 &   128.0 & 0.414 &    - & 2236.38 & 654.87 &0.024\\
359.873740 & -22.608472 & 114806580 & 17.224 & 0.001 & 16.642 &  0.007 &           1.187 &               - & 4942.0 &   124.0 & 0.757 &    - & 2582.02 & 641.74 &0.024\\
359.842153 & -21.467468 & 114806358 & 17.377 & 0.001 & 16.819 &  0.007 &           1.503 &               - & 5074.0 &   126.0 & 1.350 &    - & 3746.11 & 911.14 &0.025\\
359.828501 & -22.763815 & 114806606 & 17.480 & 0.001 & 16.963 &  0.007 &           0.786 &               - & 5275.0 &   127.0 & 0.431 &    - & 2262.61 & 656.12 &0.020\\
359.806848 & -19.078055 & 114805878 & 17.357 & 0.001 & 16.847 &  0.007 &           1.352 &               - & 5302.0 &   126.0 & 1.301 &    - & 3753.48 & 864.69 &0.019\\
 \hline \noalign {}
\hline \noalign {\vspace{-6mm}}\label{tbl:stars}
\end{tabular}
}
\\
{\raggedright
\\$^a$ \citet{Stassun2019AJ....158..138S}, $^b$ \citet{DR2} \\
{Note}: Table outlining the stellar parameters for the first 10 stars of our target list. The full table of targets and stellar parameters is available online in a machine readable format. The reduced light curves are available on Zenodo under an open-source 
Creative Commons Attribution license: \dataset[doi:10.5281/zenodo.18476061]{https://doi.org/10.5281/zenodo.18476061}.\par}

\end{table*}

\subsection{\label{sec:candidatesignal}Signal Identification and Initial Vetting}

We determined the presence of any periodic signals in the reduced light curves for each of our target stars by generating Box Least Squares (BLS) periodograms \citep{Kovacs2002A&A...391..369K} that sampled periods of 0.6-10 days. We chose 10 days as our period upper bound following the definition of hot Jupiters described in \citet{Dawson2018ARA&A..56..175D}, which allows for at least 2 transit events per TESS sector. We sample across an array of 25,000 periods within our period range of 0.6-10 days after testing on known TESS exoplanets and finding that increasing sampling past this value did not lead to further increases in BLS power. We identified all BLS power peaks that exceeded the median noise of the periodogram by 3$\sigma$, and calculated a BLS model for each peak value which included an estimate of the period, duration, and depth of each candidate transit signal. Using the estimated depth of each candidate signal we estimated the transiting object radius using the BLS transit model depth to determine whether it was consistent with a planet. {{We chose a radius upper limit of 2 R$_{Jup}$ for signals to be flagged as planetary in nature. While there are hot jupiters such as TOI-1408 b \citep{Korth2024ApJ...971L..28K} which have reported radii in the literature of $>$ 2 R$_{Jup}$, planets with these large radii are rare, and objects larger than 2 R$_{Jup}$ tend to more often be stellar in nature.}} 
{{For a signal to be significant, we required our pipeline to select only signals with an estimated planetary radius upper limit of 2 R$_{Jup}$ and a BLS peak that was 3$\sigma$ above the median value found in our BLS periodogram. To determine this we used our pipeline to calculate the median BLS peak value, and then produced a table that listed the standard deviation of each peak relative to the calculated median value. Our pipeline then flagged only the signals which were found to exceed our chosen 3$\sigma$ threshold. We selected 3$\sigma$ as our cutoff after performing a preliminary series of injection recovery tests where we found that this cutoff provided us with the lowest number of reported false positive signals while not rejecting any of our injected signals. Using this 3$\sigma$ cutoff, our initial sample of 15,552 light curves was reduced down to a list of 1,968. We vetted these remaining light curves “by eye”, where we rejected all but 16 candidate signals for having features such as sinusoidal shapes that were more consistent with variable stars, and artificial signals that appeared due to the removal of high scatter data at the edges of a given sector. We then rejected any light curves where transit features did not consistently appear across multisector data or where the transit depth was not consistent across multisector data. Our rejection of those signals left us with a single candidate that we were unable to reject using any of our previous vetting methods. 
}}

\section{\label{sec:candidates}Results}

\subsection{\label{sec:vettingcandidates}Vetting of the Strongest Remaining Candidate}

We identified one candidate signal in our target list with a calculated radius of 1.76 R$_{Jup}$ and period of 7.21 days orbiting the T$_{mag}$ = 16.838 star TIC 92223525. This target star's light curve was reduced with \texttt{eleanor} and resulted in a normalized median absolute deviation value of 0.0039.

\begin{figure*}[t!]
\center
    \subfloat{\includegraphics[width=12cm]{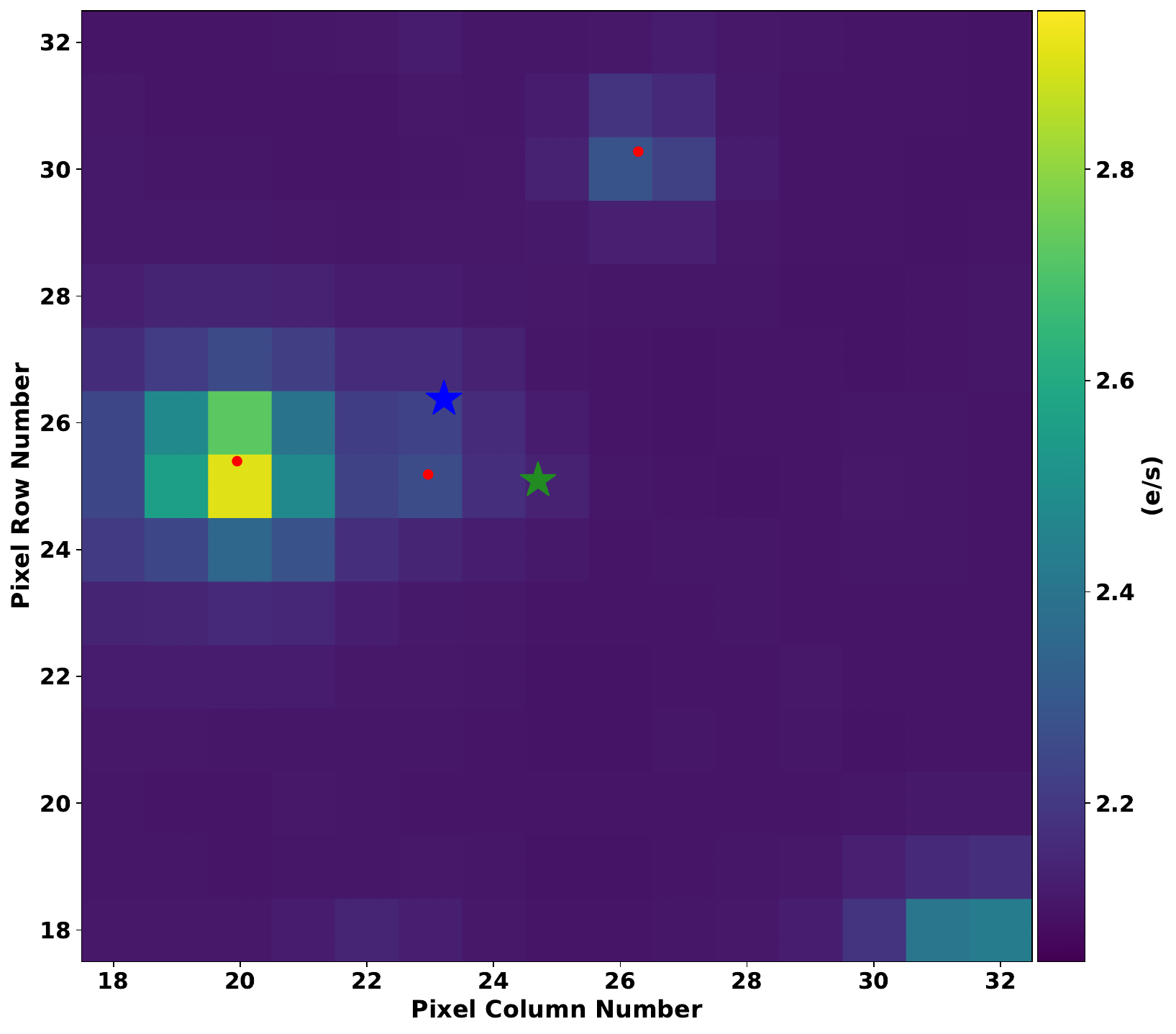}

    }
    \caption{A 15x15 Full Frame Image cutout centered on the target star TIC 92223525 (green star). The red dots represent the neighboring stars in the Gaia DR3 catalog with T$_{mag}$$<$17, and the blue star represent the neighboring star TIC 92223526 we identified as the source of the observed candidate transit signal. With a separation of just $\sim$2 TESS-pixels, we estimate that the light curve of TIC 92223525 was contaminated by 11.5\% of the observed flux from the neighboring star TIC 92223526 which we believe to be an eclipsing binary.}
    \label{fig:onsky}
\end{figure*}

\begin{figure*}[t!]
    \subfloat[]{\includegraphics[width=.49\linewidth]{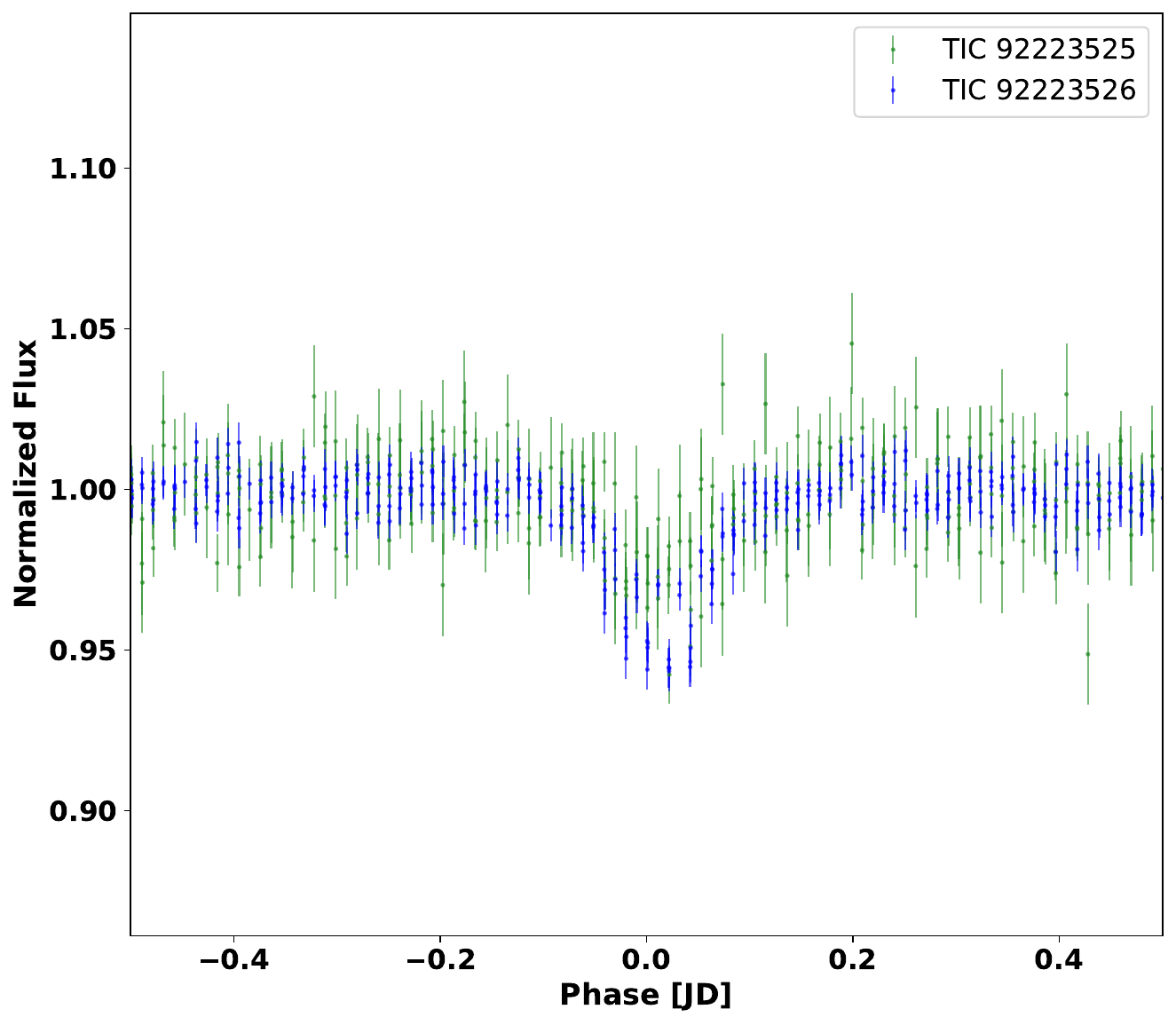}

    }
    \subfloat[]{\includegraphics[width=.49\linewidth]{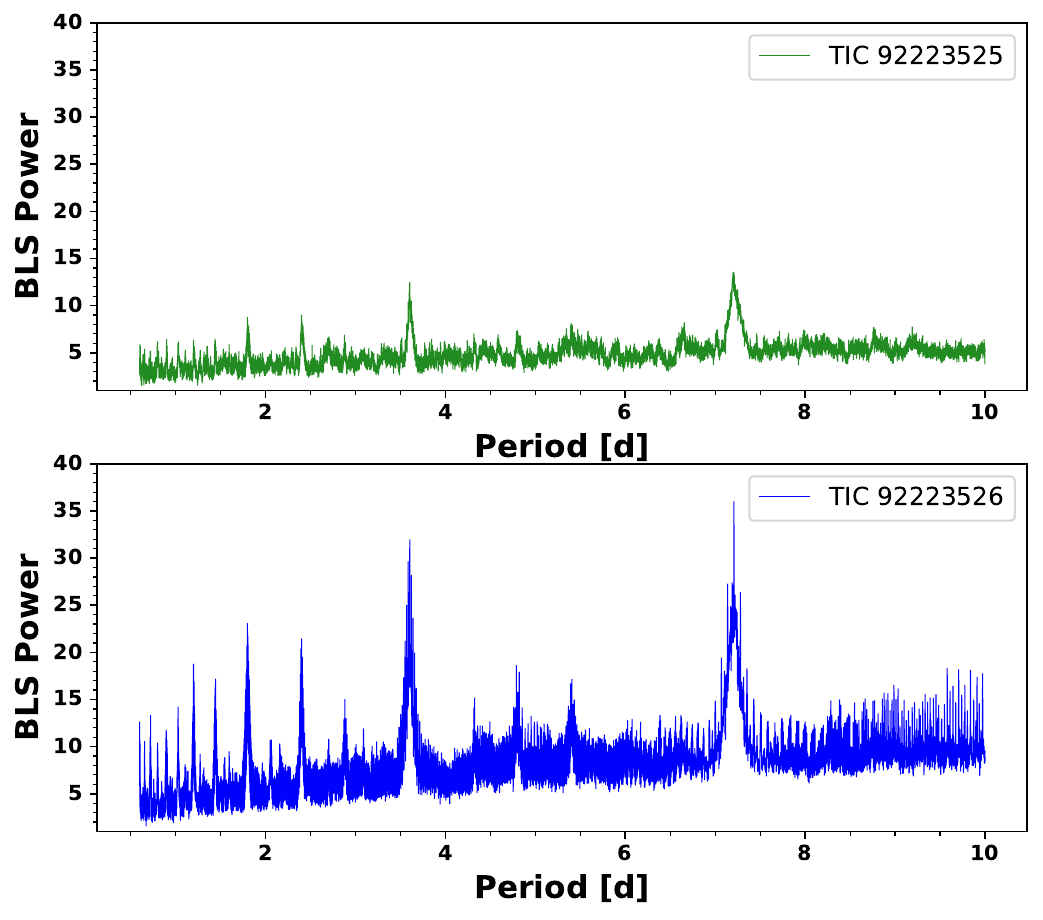}

    }\\
    \caption{Panel (a) shows the folded light curve of our candidate star TIC 92223525 overlaid with the folded light curve of the neighboring star TIC 92223526 identified as the source of the signal (blue star in Figure \ref{fig:onsky}). Each plotted color matches the star color from Figure \ref{fig:onsky}, and each dot/error bar represents the light curve data at a binned cadence of 30 minutes. In panel (b) we compare the BLS periodogram results for each star and show that the periodogram of TIC 92223526 has a much stronger power at the candidate period.}
    \label{fig:foldedlc}
\end{figure*}

To vet this signal against contamination of our aperture from a nearby star, we produce light curves for each of the surrounding stars in the TESS Full Frame Images and search for periodic signals with similar periods and mid-transit times. In Figure \ref{fig:onsky} we show TESS Full Frame Image cutouts with our target star in green, and all neighboring stars with T$_{mag}$$<$17 shown in red. We found that the light curve of the T$_{mag}$ = 15.646 star TIC 92223526 produced a stronger BLS periodogram power at the candidate period. {{The neighboring star TIC 92223526 has a stellar radius of 1.08 R$_{\odot}$ and T$_{eff}$ of 5970 K. The observed transit depth is representative of a companion with a radius of 2.4 R$_{Jup}$, larger than the planetary radius cutoff in our survey.}} In Figure \ref{fig:foldedlc} we compare our candidate with the neighboring star most likely to be the origin of the signal by showing the folded light curve and BLS periodogram of our candidate star in green, and the neighboring star in blue. 

While both \texttt{eleanor} and \texttt{TGLC} make efforts to minimize contaminating flux in crowded regions, our results show that this removal was not complete for this dim target in our survey. {{The light curve of TIC 92223525 was measured to have a median flux value of 24.53 e$^{-}$/s, and the light curve of TIC 92223526 was measured to have a median flux value of 105.81 e$^{-}$/s.}} 
We estimate our target star was contaminated by 11.5\% of the flux observed from TIC 92223526, and we therefore rejected the candidate signal as a false positive originating from flux contamination from a nearby eclipsing binary. 

\subsection{\label{sec:recovery}Determining Survey Completeness}

We determined the completeness of our survey through injection-recovery testing. For each of our targets stars we used the transit modeling package \texttt{batman} \citep{kreidberg2015} to simulate 10,000 injected planet signals with properties drawn from a uniform random combination of planet radii ranging from 1-2 R$_{Jup}$ and orbital periods ranging from 0.6-10 days. We selected these transiting planet parameters to encompass a wide range of radius values commonly associated with hot Jupiters \citep{Dawson2018ARA&A..56..175D} at transit mid-times that were randomly selected within the observation window for each target. {{We set the eccentricity of these injected planets to "0" given that the low orbital periods of hot Jupiters often result in tidal circularization of their orbits. }}
 
{{After the injection of these signals, we performed the same cleaning procedure and BLS periodogram analysis used on the light curves during our planet search (see Section \ref{sec:candidatesignal}). This included the removal of signals with a radius of 2 R$_{Jup}$ as well as rejecting all BLS periodogram signals with signal-to-noise values below 3$\sigma$. If any of these periods matched the injected signal's period to within 5\%, we classified them as a successful recovery.}}
 
\begin{figure*}[htp]
\centering
\includegraphics[width=16cm]{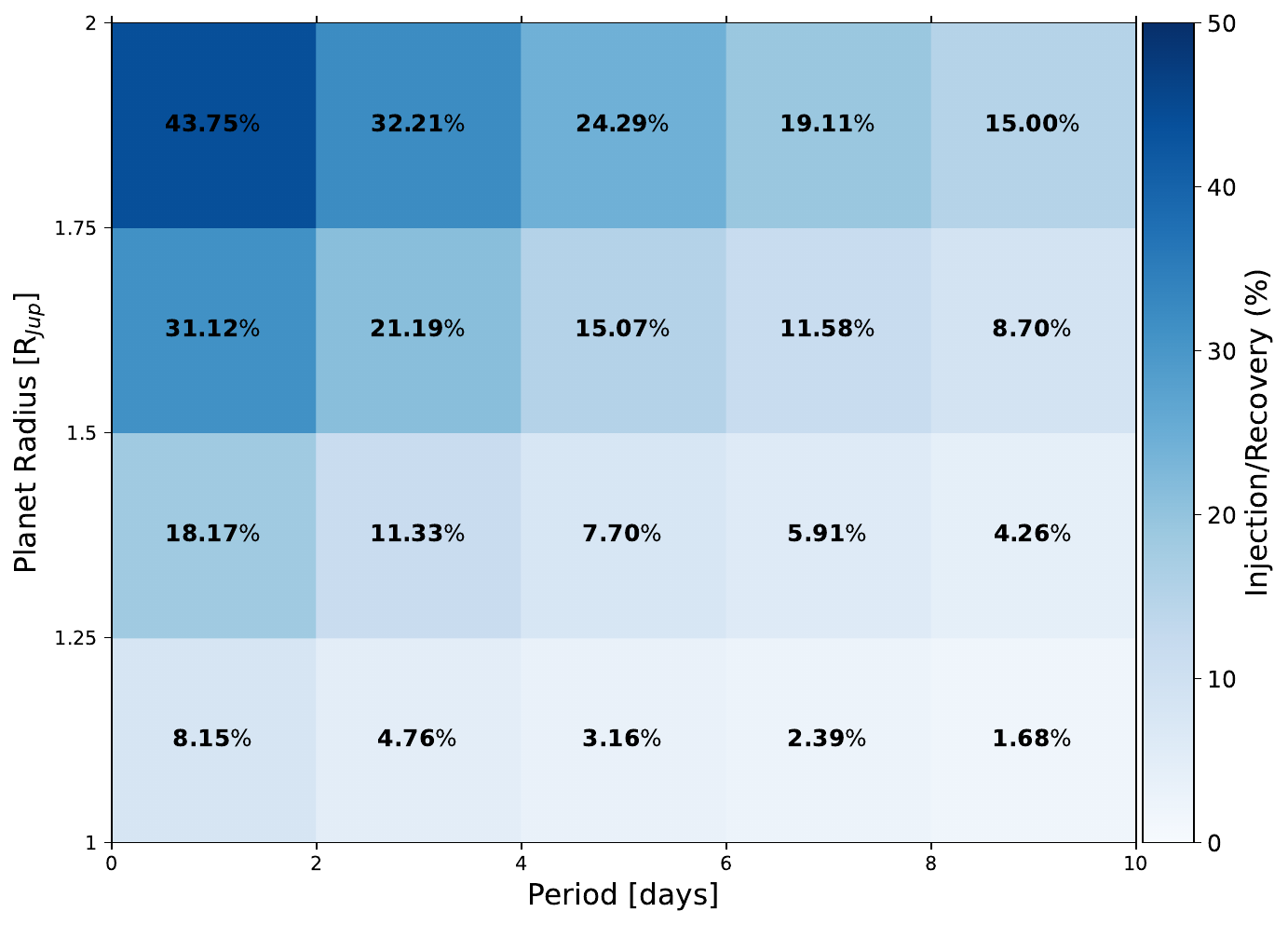}
\caption{Injection recovery rates for 2,481 targets from our survey of 15,176 stars in the Sagittarius dwarf galaxy stream. For each star we performed 10,000 recovery tests, where each injection included a planet with parameters that were drawn from uniform random samples of radii between 1-2 R$_{Jup}$ and orbital periods less than 10 days. Each box represents a bin of 2 days on the X-axis and 0.25 R$_{Jup}$ on the Y-axis. The center of each box is labeled with the percent of signals recovered from within that bin calculated as (total successful recoveries/total injections)*100. The overall color gradient reflects the recovery percentage value.} \label{fig:inject}
\end{figure*}

In Figure \ref{fig:inject} we present the combined results of the injection recovery tests for 2,481 target stars from our survey. Each box within the grid represents a bin size of 2 days in period on the X-axis, and 0.2 R$_{Jup}$ in planet radii on the Y-axis. To calculate the recovery rate percentage inside each box we took the number of successfully recovered planets within the bin range of period and radius and divided by the total number of injected signals within each bin range. At periods of less than 2 days we can recover 43.75\% of 1.8-2 R$_{Jup}$ planets, and 8.15\% of 1-1.2 R$_{Jup}$ planets. For planets orbiting at the longest hot Jupiters orbital periods of 8-10 days, we recover 15.00\% of 1.8-2 R$_{Jup}$ planets and 1.685\% of 1-1.2 R$_{Jup}$ planets.

\subsection{\label{sec:ocurr}Upper Limits on the Extragalactic hot Jupiter Occurrence Rate of the Sagittarius Dwarf Galaxy}

We detected no significant planet signals in our survey of 15,176 stars, and calculated the occurrence rate upper limit for extragalactic hot Jupiters by combining the results of our injection-recovery testing (see Section \ref{sec:recovery}) with the equation:

\begin{equation}
  P_{null} = \prod_{i=1}^{n} 1 - \eta d_i,
\label{occureq}
\end{equation}
where $\eta$ is the occurrence rate upper limit, $d_i$ is the detection efficiency, and $P_{null}$ is the confidence level of the upper limit $\eta$. The detection efficiency is related to both the recovery efficiencies from our injection recovery testing and the geometric transit probability defined as:

\begin{equation}
d_i = P_{recovery} \times P_{transit},
\label{eq:di}
\end{equation}
where P$_{recovery}$ is the injection recovery percent per bin, and P$_{transit}$ is the geometric transit probability per star within a given bin. To determine the geometric transit probability we use the equation:

\begin{equation}
P_{transit} = \frac{R_{p,med}\times R_{*}}{a_{p,med}},
\label{tranprob}
\end{equation}
where $R_{*}$ is the stellar radius, and $R_{p,med}$ and $a_{p,med}$ are the median planet radius and orbital semimajor axis for the chosen bin. We selected the value of P$_{recovery}$ using the corresponding grid value from the injection recovery results.

\begin{figure*}[htp]
  \centering
  \begin{tabular}{@{}c@{}}
    \includegraphics[width=0.75\textwidth]{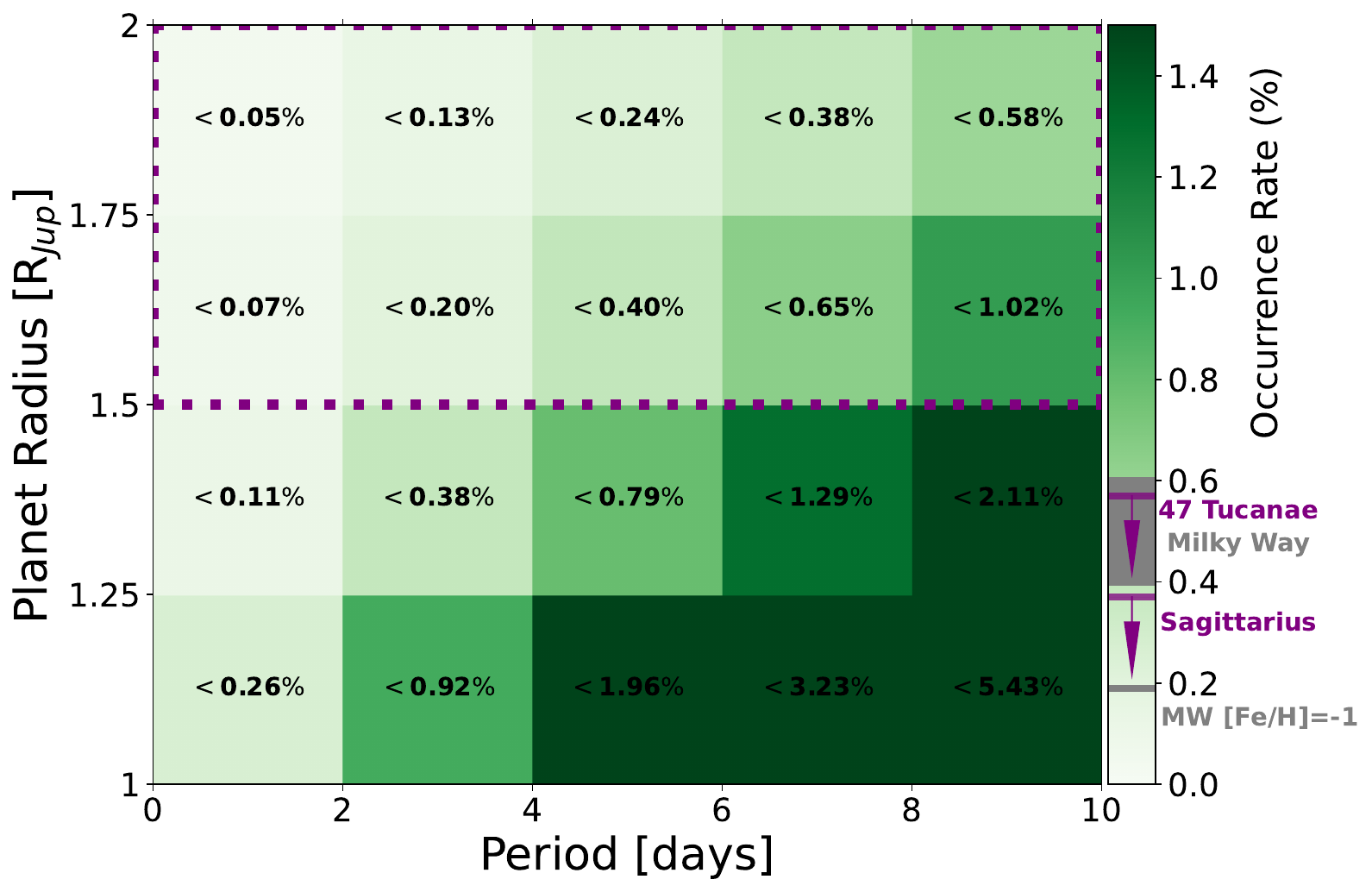} \\
    \small (a)
  \end{tabular}

  \begin{tabular}{@{}c@{}}
    \includegraphics[width=0.75\textwidth]{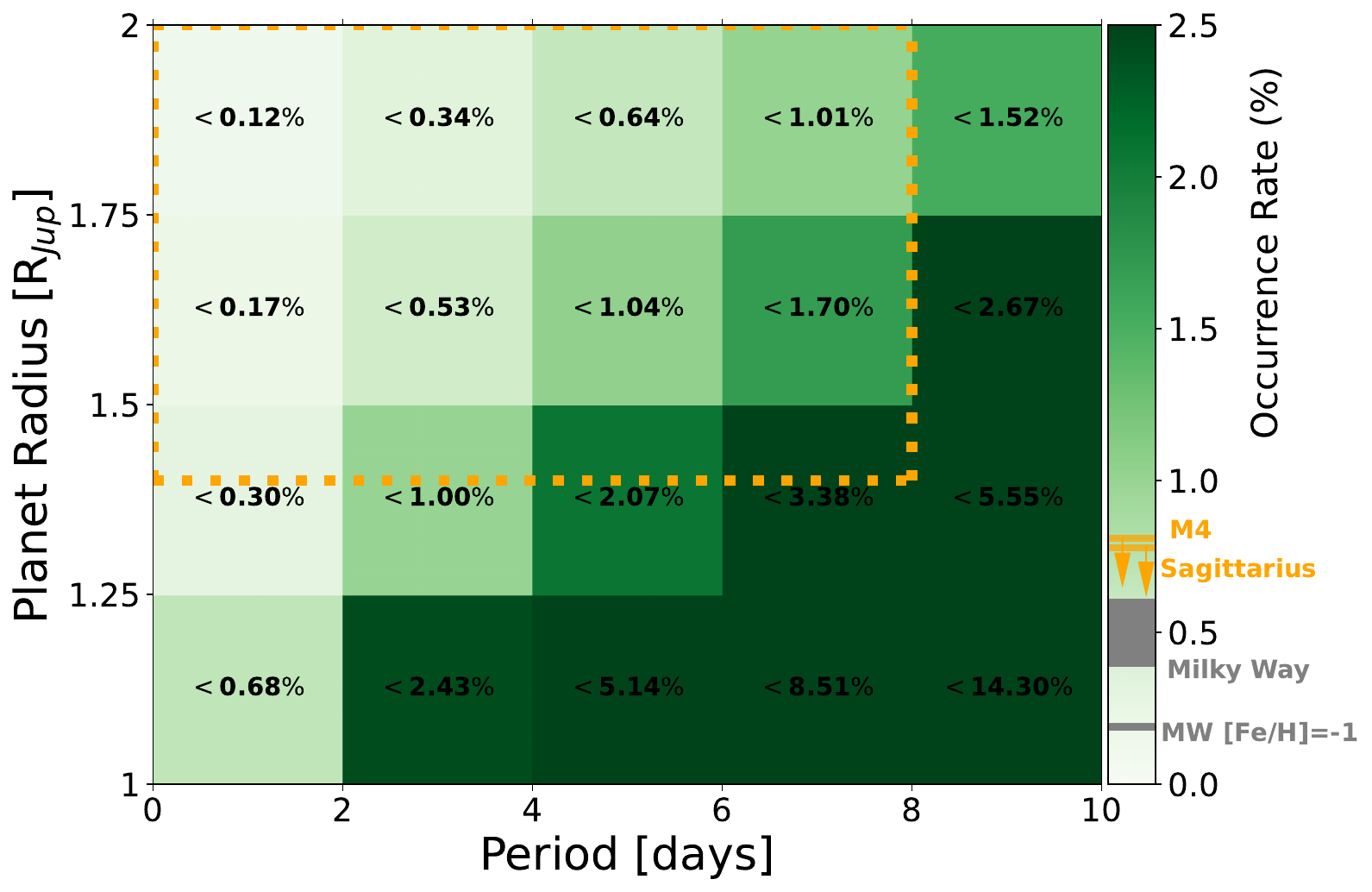} \\
    \small (b) 
  \end{tabular}

  \caption{The first upper limit on the hot Jupiter occurrence rate in the Sagittarius dwarf galaxy across a radius range of 1-2 R$_{Jup}$ with orbital periods $<$ 10 days. Occurrence rates at the 68\% and 95\% confidence levels are shown in panels (a) and (b) respectively, and were calculated using Equation \ref{occureq}. The values at the center of each box represent the occurrence rate upper limit for that bin in planetary parameters shown as a percent. For 1-2 R$_{Jup}$ hot Jupiters with periods of 0.6-10 days we calculate an average 1$\sigma$ occurrence rate upper limit of $<$1.01\%. In panel (a) we compare our results to the 1-$\sigma$ occurrence rate upper limit of $<$0.57\% calculated by \citet{Crisp2024arXiv241209705C} for the globular clusters 47 Tucanae within the parameter space of 1.5-2 R$_{Jup}$ and $<$10 day periods outlined by the dotted purple box. Within the same parameter space we find a 1-$\sigma$ occurrence rate upper limit value of $<$0.37\% for the Sagittarius dwarf galaxy stream. In panel (b) we compare our results to the 2-$\sigma$ occurrence rate upper limit of $<$0.81\% calculated by \citet{Wallace2020AJ....159..106W} for the globular cluster M4 within the parameter space of 1.4-2 R$_{Jup}$ and $<$8 day periods outlined by the dotted orange box. Within the same parameter space we find a 2-$\sigma$ occurrence rate upper limit value of $<$0.78\% for the Sagittarius dwarf galaxy stream. {{The color bars of each panel are labeled with the calculated upper limits for each region within the highlighted parameter space. The gray box on each color bar represents the range of 0.4-0.6\% for hot Jupiter occurrence rate seen in the Milky Way, and the gray bar labeled ``MW [Fe/H]=-1" represents the Milky Way hot Jupiter occurrence rate scaled to a [Fe/H]=-1 representative of the low metallicity environment of the Sagittarius dwarf galaxy stream.}}}
  \label{fig:occur}
\end{figure*}

Using Equation \ref{occureq}, we calculated occurrence rate upper limit measures for $\eta$ in each bin of orbital period and planet radius. We began with an occurrence rate guess of 80\% which is much larger than the expected occurrence rate. The value of $\eta$ is then iteratively reduced until we determined the calculated value of $P_{null}$ in Equation \ref{occureq} that is consistent with the desired confidence level. The final occurrence rate upper limits at both the 68\% and 95\% confidence level are presented in Figure \ref{fig:occur}. We calculate an average occurrence rate upper limit of $<$1.01\% across our entire tested parameter space of 1-2 R$_{Jup}$ hot Jupiters with orbital periods $<$ 10 days at the 68\% confidence level, and find an occurrence rate upper limit of $<$0.31\% for radii of 1-2 R$_{Jup}$ with orbital periods $<$4 days. The highlighted purple and orange regions of each panel in Figure \ref{fig:occur} represent the parameter space we used for our comparison to the Milky Way globular clusters M4 and 47 Tuc in Section \ref{sec:globularclusters}. The color bar of each panel is labeled with the upper limit values for the Sagittarius dwarf galaxy and the corresponding Milky Way globular cluster upper limit. Our calculated hot Jupiter occurrence rate upper limits represent the first time a large transit survey has been performed on an individual extraglactic-origin stream. With a target list magnitude limit of T$_{mag}$$<$17 and stellar light curves that approach the noise floor of the TESS Full Frame Images, our survey performs at the most extreme limits of current TESS observations. 

\section{\label{sec:discussion}Discussion}

\subsection{\label{sec:globularclusters}Comparison to Milky Way Stellar Clusters}

{{Our measured occurrence rate upper limit of $<$1.01\% for hot Jupiters with radii of 1-2 R$_{jup}$ and periods of 0.6-10 days is higher than the observed occurrence rates of 0.4-0.6\% seen by Kepler across the Milky Way \citep{Howard2012ApJS..201...15H,Fressin2013ApJ...766...81F, Mulders2015ApJ...798..112M, Petigura2018AJ....155...89P, Kunimoto2020AJ....159..248K}, or the 0.98 $\pm$ 0.36 \% for G stars seen by TESS \citep{Beleznay2022}. However, both of these samples include stars with higher metallicities and younger ages which make direct comparisons more difficult. We instead compare our value to what has been seen in other Milky Way halo populations.}}

The Milky Way's halo consists of of stellar streams and globular clusters that result from galaxy mergers that have occurred throughout the Milky Way's lifetime. We compare our occurrence rate upper limits to the upper limits seen for globular clusters in the Milky Way halo to put our results in the context of other merger related environments. 

\citet{Crisp2024arXiv241209705C} recently conducted a transit survey of 19,930 stars in the [Fe/H]=-0.78 globular cluster 47 Tucanae using the Dark Energy Camera. For the 1.5-2 R$_{Jup}$ planets our survey was most sensitive to, their survey found an average occurrence rate upper limit of $<$0.57\% at the 68\% confidence level, which is higher than the upper limit of $<$0.37\% we found across the same parameter space and confidence level in the Sagittarius dwarf galaxy stream (see purple box and labels in panel (a) of Figure \ref{fig:occur}). 

\citet{Wallace2020AJ....159..106W} conducted a transit survey of the [Fe/H]$\simeq$-1.2 globular cluster M4 (NGC 6121) using observations of 4,554 stars from the Kepler K2 mission with magnitudes of G$<$19. For planets with radii of 1.4-2 R$_{Jup}$ and periods of 0-8 days (see orange dashed box in Figure \ref{fig:occur}), they report an occurrence rate upper limit of $<$0.81\% at the 95\% confidence level. Within the same parameter space and confidence level we find a slightly lower occurrence rate upper limit of $<$0.78\% for the Sagittarius dwarf galaxy stream (see orange box and color bar labels in panel (b) of Figure \ref{fig:occur}).

\begin{table}[t]
\begin{center}
\caption[]{\em{Upper Limit Occurrence Rate Values for Sagittarius vs Milky Way Globular Clusters}}
\begin{tabular}{ccccc}
\hline \noalign {}
\hline \noalign {\smallskip}
\hline \noalign {\smallskip}
Survey & Planet Radii & Periods & $\eta$ & Confidence Level\\
       & (R$_{Jup}$)  & (days) &   (\%)  &       (\%)      \\
\hline \noalign {\smallskip}
47 Tucanae  & 1.5-2    &  $<$10d & $<$0.57\% &    68\%     \\
M4          & 1.4-2    &  1-8d   & $<$0.81\% &    95\%     \\
Sagittarius & 1.5-2    &  $<$10d & $<$0.37\% &    68\%     \\
            & 1.4-2    &  1-8d   & $<$0.78\% &    95\%     \\
 \hline \noalign {}
\hline \noalign {\vspace{-6mm}}\label{tbl:occur}
\end{tabular}
\end{center}
{\raggedright {Note}: Comparison between our occurrence rate upper limits calculated for the Sagittarius dwarf galaxy stream and the upper limits found in the globular clusters M4 \citep{Wallace2020AJ....159..106W} and 47 Tucanae \citep{Crisp2024arXiv241209705C} within the listed parameter spaces. The listed planetary radii, periods, and confidence level values used to determine the upper limits for the Sagittarius dwarf galaxy stream were selected to match the values used by \citet{Crisp2024arXiv241209705C} and \citet{Wallace2020AJ....159..106W} in their calculations.\par}
\end{table}

\subsection{\label{sec:etallicityeffect} Lower Metallicity}

{{Compared to the average metallicity of [Fe/H] = -0.2 measured for the Kepler sample \citep{Dong2014ApJ...789L...3D}, the Sagittarius dwarf galaxy stream contains a stellar population which is much more metal poor ([Fe/H] = -2.3-0.0 \citep{minelli2023}). RV and transit surveys across the varying environments of the Milky Way have indicated that giant planets are more frequent around higher metallicity stars \citep{fischer2005,Boley2021AJ....162...85B,Thorngren2016ApJ...831...64T}. That trend appears to continue down to the metal-poor regime (-2.0$\leq$ [Fe/H] $\leq$ -0.6) where a mean 1$\sigma$ upper limit was found to be 0.18\% for planet radii of 0.8-2 R$_J$ and periods of 0.5-10 days. Hot Jupiter formation therefore appears to be strongly inhibited at and below this point \citep{Boley2021AJ....162...85B}.}} It should therefore be expected that the low metallicity environment of the Sagittarius dwarf galaxy experiences a low hot Jupiter occurrence rate. The observed trend between metallicity and giant planets was defined by \citet{fischer2005} using the power law relation:  

\begin{equation}
f_{p} \propto  10^{\beta[Fe/H]}
\label{metalprob}
\end{equation}

where $f_{p}$ is the fraction of stars with planets, [Fe/H] is the representative metallicity value, and $\beta$ is the index of the power law. We select the $\beta$ value of 0.71$^{+0.56}_{-0.34}$ which was derived by \cite{Osborn2020} for hot Jupiters.

We can use Equation \ref{metalprob} to determine whether the hot Jupiter occurrence rate upper limit for the Sagittarius dwarf galaxy stream matches what would be expected based on low metallicity Milky Way environments. We calculate that this expected occurrence rate $f_{p}$ would be $\sim$ 0.19\% at the average metallicity of [Fe/H] = $\sim$-1 seen in the Sagittarius dwarf galaxy stream. Our overall 1$\sigma$ average occurrence rate upper limit of 1.01\% is larger than this predicted value. 

Following the mass-metallicity relation observed for galaxies \citep{Kirby2013}, the other Milky Way satellite dwarf galaxies we plan to survey in future work have low average metallicities similar to the Sagittarius dwarf galaxy stream. With the larger number of observed low metallicity extragalactic stars we will observe, we will be able to place even stronger constraints on the extragalactic hot Jupiter occurrence rate. 

We estimate the number of stars needed to detect a planet at the metallicity of the Sagittarius dwarf galaxy with the equation:

\begin{equation}
\eta_{extragalactic} \geq \frac{1}{\eta_{HJ}*P_{transit}*P_{SGR}*k_{[Fe/H]}}, 
\label{sampleincrease}
\end{equation}

{{using the methodology from \citet{Yoshida2022AJ....164..119Y} where $\eta_{HJ}$ is the Milky Way hot Jupiter occurrence rate range (0.4-0.6\% \citep{Howard2012ApJS..201...15H,Fressin2013ApJ...766...81F, Mulders2015ApJ...798..112M, Petigura2018AJ....155...89P, Kunimoto2020AJ....159..248K}), $P_{transit}$ is the average transit probability for our hot Jupiters (we assume 0.09 for our sample using Equation \ref{tranprob} averaged across our period range of $<$10 days), and $k_{[Fe/H]}$ is the factor of metallicity on planet occurrence (0.19 from Equation \ref{metalprob} assuming an average [Fe/H] of -1). While \citet{Yoshida2022AJ....164..119Y} included a kinematic mixing factor of P$_{KM}$ to represent that $\sim$50\% of their stars were kinematically mixed to their present position from within the Milky Way, we replace this factor with the average probability of Sagittarius dwarf galaxy membership for our sample, $P_{SGR}$, as our sample was explicitly constructed to include stars with membership probabilities $>$50\%. We estimate that with a perfect detection efficiency this would require $\eta_{extragalactic}$ $\geq$ 11,467 target stars based on the observed range in Kepler hot Jupiter occurrence rates to either detect a planet of extragalactic origin, or determine that the expected trend between metallicity and occurrence rate cannot fully explain the hot Jupiter occurrence rate of extragalactic environments. In order to determine the number of stars needed for a survey with the same detection efficiency as ours, we divided this number of stars by the average recovery rate of 14.48\% seen across our measured parameter space and determined it would require a survey size of $\geq$79,189 stars. A future search of additional extragalactic streams will approach this value.}}

\subsection{\label{sec:binary}Higher Stellar Binarity}

It is predicted that the binary population of a galaxy is strongly dependent on the star formation rate (SFR), and that the low SFR seen in dwarf galaxies are expected to exhibit a significantly higher binary fraction than the Milky Way \citep{Marks2011MNRAS.417.1702M}. Observations of the Sagittarius dwarf galaxy have shown this to be true with a fraction of RV variables that is a factor of $\sim$2 higher than that of the Milky Way, suggesting that its fraction of close binaries is therefore intrinsically higher \citep{Bonidie2022ApJ...933L..18B}. 

The effect of binarity on planet formation has been investigated across the Milky Way. Using a combined multiplicity/disk census of multiple star forming regions, \citet{Kraus2012ApJ...745...19K} find that $\sim$2/3 of all close ($<$40 au) binaries disperse their disks within $\sim$1 Myr of their formation due to the tidal influence of the companion, and $\sim$80\%–90\% of wide binaries and single stars retain their disks for at least $\sim$2–3 Myr. Furthermore \citet{Kraus2016AJ....152....8K} present high resolution imaging of 382 Kepler Objects of
Interest (KOIs) and find that the influence of a binary companion should suppress the formation of planetary systems in a fifth of solar type stars. \citet{Moe2021MNRAS.507.3593M} combine radial velocity and imaging surveys to shown that stars with a close ($<$ 1 au) stellar binary companion can fully suppress the formation of circumstellar planets, while binaries with more distant companions ($>$ 200 au) appear to have a negligible effects on the formation of planets close to their host star. For smaller planets a close binary results in a lower occurrence rate than is observed around single stars. However, binary interactions may have positive effects on the creation of higher mass planets. For close binaries, the merging of two low mass stars may lead to the formation of hot Jupiters in the massive, compact disk left after the merger \citep{Martin2011A&A...535A..50M}. 

In systems with binaries on wider orbits, the stellar companion may provide the dynamical effects needed to transition a cold Jupiter into a hot Jupiter. According to migration theories for hot Jupiter formation, a Jupiter mass planet is first formed in the outer regions of a protoplanetary disk, and then migrates inward to its present day close-in orbits. A significant fraction of observed hot Jupiters ($\sim$10\%) can be explained by migration induced by a binary companion which is highly inclined to the orbit of the planet \citep{Wu2007ApJ...670..820W}. The binary fraction of the Sagittarius dwarf galaxy stream is $\sim$2x higher than is seen in the Milky Way, and this increased binarity may lead to a planet forming environment where Jupiter sized planets can more readily migrate from the outer to inner disk through dynamical interactions. 

The increased binary fraction of the Sagittarius dwarf galaxy stream may also indicate that it experienced a different star formation history than the Milky Way. The distribution of stellar masses within a star forming region is described by the stellar initial mass function (IMF), and observations of more high-mass star formation occurring in high redshift galaxies than is seen in lower redshift galaxies indicate that the IMF evolves over cosmic time on a galactic scale \citep{Dave2008,jerabkova2018}. Although \citet{mann15} found no evidence for a disk mass dependence on the projected separation from the massive star in NGC 2024, \citet{Winter2022EPJP..137.1132W} show that the UV radiation of OB stars can directly impact planet formation by truncating the protoplanetary disks like the proplyds in the Orion Nebular Cluster through external photoevaporation. This truncation would reduce the already low protoplanetary disk dust masses of low mass stars \citep{andrews13} to the point where they can no longer create the massive cores required to produce giant planets. An IMF which produced a large number of high mass stars in the Sagittarius dwarf galaxy would have therefore made the formation of hot Jupiters more difficult than would be expected for a region with a high binary fraction.

\subsection{\label{sec:stellarage}Older Stellar Population}

To be detected in a transit survey a hot Jupiter must not only form within a given star forming environment, but must survive long enough to be observed. Hot Jupiters with periods of $\leq$1 day experience orbital decay times of $\leq$100 Myrs for main sequence stars with stellar masses $\leq$1.0 M$_{\odot}$ \citep{Weinberg2024ApJ...960...50W}, and rapid orbital decay on timescales shorter than the main sequence lifetimes of solar-type stars may explain their lack of giant planets observed with orbits $\leq$2 days \citep{Essick2016ApJ...816...18E}. As these stars evolve off of the main sequence and onto the red giant branch the increase in their stellar radius leads to the orbital decay of planets with orbits of $a$/R$_{*}$ $\sim$ 2-3, resulting in planetary engulfment \citep{Villaver2014ApJ...794....3V}. 

The mix of young, intermediate, and old populations of stars in the Sagittarius dwarf galaxy stream \citep{ramos22} means that the orbital decay of hot Jupiters has likely occurred for some of the old and intermediate aged stars in the stream. If our target list is biased towards more older stars than were observed with Kepler in the Milky Way, the occurrence rate of hot Jupiters in the Sagittarius dwarf galaxy would be expected to be lower due to orbital decay of hot Jupiters over time for the $\leq$ 2 R$_{\odot}$ stars in our target list.

\subsection{\label{sec:stellardensity}Effects from Low Stellar Density}

Close stellar flybys of neighboring stars in regions of high stellar density can induce the inward migration of Jupiter sized planets. Even after accounting for metallicity differences, hot Jupiters are observed to occur more frequently in regions of denser stellar clustering due to these dynamical interactions \citep{Winter2020}. If disk migration induced by high stellar densities plays a large role in the formation of hot Jupiters, the low stellar density of the Sagittarius dwarf galaxy \citep{Bellazzini1999MNRAS.307..619B} may therefore help explain its low occurrence rate. 

While high stellar densities appear to induce the disk migration needed to transform a giant planet from a cold Jupiter to a hot Jupiter, low stellar density environments do not appear to inhibit cold Jupiter formation on orbits longer than 1 year \citep{Gratton2023NatCo..14.6232G}. If low stellar density leads to a bias towards more cold Jupiters than hot Jupiters, then the low occurrence rate of hot Jupiters seen in the Sagittarius dwarf galaxy stream may instead reflect a stellar population that follows a Milky Way-like trend with metallicity that appears further suppressed due to a larger number of non-migrated cold Jupiters. The detection of cold Jupiters in the Sagittarius dwarf galaxy stream rather than hot Jupiters would provide evidence to this theory. Furthermore, the relationship between cold Jupiters and super Earths would predict that $\sim$90\% of those cold Jupiters would be paired with super Earths with orbital radii of $<$1 au \citep{Zhu2018AJ....156...92Z}.

\subsection{\label{sec:recentpassage}Dynamical Perturbations Caused by the Most Recent Passage of the Sagittarius Dwarf Galaxy}

The release of Gaia DR2 led to the discovery of a spiral-shaped distribution in the vertical position–velocity (Z–V$_{Z}$) phase space for stars in the Galactic disk \citep{Antoja2018Natur.561..360A}, and N-body simulations have shown that this feature likely results from the last pericentric passage of the Sagittarius dwarf galaxy 500-800 Myr ago \citep{Laporte2019MNRAS.485.3134L}. For the Milky Way planetary systems found in this spiral, the ratio of hot to cold Jupiters is an order of magnitude higher than the ratio seen for field stars outside this phase space feature. \citet{Kruijssen2021arXiv210906182K} suggest that this increase is likely caused by the dynamical perturbations of existing planetary systems by stellar encounters or galactic tides during the last passage of the Sagittarius dwarf galaxy. If the same dynamic perturbations that led to the increase in hot Jupiters for the Milky Way were experienced by the passing Sagittarius dwarf galaxy, it would be expected that the Sagittarius dwarf galaxy also experienced an increase in its hot Jupiter occurrence rate.

\section{\label{sec:conclusion} Conclusions and Future Work}

We use the light curve reduction packages \texttt{eleanor} and \texttt{TGLC} to produce calibrated light curves for 15,176 stars in the Sagittarius dwarf galaxy stream and conduct the first transit survey of an extragalactic origin stellar stream. We identified 1 candidate planet signal which we rejected due to its periodic signal being better attributed to aperture contamination from a nearby star found in the TESS Full Frame Images. We use a series of injection-recovery tests to determine our completeness for detecting hot Jupiters with radii ranging from 1-2 R$_{Jup}$ and orbital periods of 0.6-10 days, and calculated the first occurrence rate upper limit for hot Jupiters in the Sagittarius dwarf galaxy stream. We calculate a 1$\sigma$ average upper limit of $<$1.01\% across this parameter space. {{We estimate that with a perfect detection efficiency this would require $\eta_{extragalactic}$ $\geq$ 11,467 target stars based on the observed range
in Kepler hot Jupiter occurrence rates to detect a planet of extragalactic origin. With a detection efficiency comparable to this survey this would require $\eta_{extragalactic}$ $\geq$ 79,189 target stars.}}

While the Sagittarius dwarf galaxy stream was chosen for the analysis in this work due to its relatively large population of identified member stars at T$_{mag}$$<$17, the Milky Way halo consists of many more stellar streams of extragalactic origin. These include streams such as Helmi \citep{Helmi1999Natur.402...53H,helmistream2019A&A...625A...5K}, Palomar-5 \citep{Odenkirchen2009AJ....137.3378O}, GD-1 \citep{Grillmair2006ApJ...643L..17G}, Orphan \citep{Grillmair2006ApJ...643L..17G}, Fimbulthul \citep{Ibata2019ApJ...872..152I}, and Price-Whelan 1 \citep{Price-Whelan2019ApJ...887...19P}. Future work surveying these additional streams will either result in the detection of the first extragalactic origin planet, or provide the first evidence that extragalactic environments may experience a lower hot Jupiter occurrence rate than the Milky Way.

\subsection{\label{sec:level} Software}

\texttt{NumPy} \cite{harris2020array},
\texttt{Astropy \citep{astropy:2013, astropy:2018, astropy:2022}},
\texttt{Astroquery \citep{astroquery2019}},
\texttt{IPython \citep{ipython:2007}},
\texttt{Matplotlib \citep{matplotlib:2007}},
\texttt{pandas \citep{pandas2010}},
\texttt{SciPy \citep{2020SciPy}},
\texttt{eleanor} \cite{Feinstein2019PASP..131i4502F},
\texttt{tess-point} \cite{Burke2020ascl.soft03001B},
\texttt{TessCut} \cite{Brasseur2019ascl.soft05007B},
NASA GSFC-\texttt{eleanor lite}\cite{Powell2022RNAAS...6..111P},
\texttt{TGLC} \cite{han2023},
\texttt{lightkurve} \cite{ 2018ascl.soft12013L}

\subsection{\label{sec:level} Acknowledgments}

This research has made use of the NASA Exoplanet Archive, which is operated by the California Institute of Technology, 
under contract with the National Aeronautics and Space Administration under the Exoplanet Exploration Program.
Specifically, it made use of the Transiting Planets Table \citep{10.26133/nea37}.
Some/all of the data presented in this article were obtained from the Mikulski Archive for Space Telescopes (MAST) 
at the Space Telescope Science Institute; the data for the {\it TESS} mission is available at MAST \citep{10.17909/fwdt-2x66}.

This work has also made use of data from the European Space Agency (ESA) mission
{\it Gaia} (\url{https://www.cosmos.esa.int/gaia}), processed by the {\it Gaia}
Data Processing and Analysis Consortium (DPAC,
\url{https://www.cosmos.esa.int/web/gaia/dpac/consortium}). Funding for the DPAC
has been provided by national institutions, in particular the institutions
participating in the {\it Gaia} Multilateral Agreement.

\facilities{TESS, Gaia, Exoplanet Archive}
\clearpage
\bibliography{new.ms}{}
\bibliographystyle{aasjournalv7}

\end{document}